\documentclass[letterpaper,twocolumn,10pt]{article}
\usepackage{usenix,epsfig,endnotes,indentfirst}
\begin{document}

\date{}

\title{\Large \bf Making Availability as a Service in the Clouds}

\author{
{\rm Pengfei Chen \ Yong Qi \ Peipei Wang \ Li Su \ Xinyi Li}\\
School of Electronic and Information Engineering \\
Xi'an Jiaotong University\\
fly.bird.sky@stu.xjtu.edu.cn
\and
} 

\maketitle

\thispagestyle{empty}

\subsection*{Abstract}
 Cloud computing has achieved great success in modern IT industry as an excellent computing paradigm
due to its flexible management and elastic resource sharing. To date, cloud computing takes an 
irrepalceable position in our socioeconomic system and influences almost every aspect of our daily
life. However, it is still in its infancy, many problems still exist.Besides the hotly-debated security problem, 
availability is also an urgent issue.With the limited power of availability mechanisms provided in present cloud platform,
we can hardly get detailed availability information of current applications such as the root causes of availability problem,mean time to failure,etc. 
Thus a new mechanism based on deep avaliability analysis is neccessary and benificial.Following the prevalent terminology `XaaS',this paper proposes a new win-win concept for cloud users and providers in term of `Availability as a Service' (abbreviated as `AaaS').The aim of `AaaS' is to provide comprehensive and aim-specific \textsl{ \textbf {runtime}} avaliabilty analysis 
services for cloud users by integrating plent of data-driven and model-driven approaches. To illustrate this concept, we realize a prototype named `EagleEye' with all features of `AaaS'.
By subscribing corresponding services in `EagleEye',cloud users could get specific availability information of their applications deployed in cloud platform. We envision this new kind of service will be merged into the cloud management mechanism in the near future. 
\section{Introduction}
 Cloud computing has become a dominant computing paradigm in modern IT world due to its essential appealing Characteristics: on-demand self-service, broad network access, resource pooling, rapid elasticity, measured service \cite{1}. Compared with the traditional computing mode, cloud computing enables service providers rapidly delivery and deploy services on the cloud platform remotely with little effort instead of setting up the hardware or software infrastructure locally. Under the attraction of benefit receiving from cloud computing, more and more service providers shift to the clouds. However, cloud computing is far from `mature', many issues still exist. Some issues are apparent with much discussion such as security and performance issues. While others may be hidden with little discussion such as availability \cite{2} and risk assessment \cite{3} issues in clouds. This paper will concentrate on the availability-related issues and propose some insights on that. Our goal is to improve the runtime troubleshooting support of present availability management mechanisms in clouds by introducing a new kind of service.

 Availability is an important metric in SLA (Service Level Agreement). And high availability is a critical reason for cloud users embracing cloud computing. However, it didn't get the desired attention when cloud computing just appeared. In 2011 the Amazon's cloud crash makes a hard strike to people's confidence in cloud computing. Since then, the availability issue in the clouds has received tremendous attention in academic community and IT industry. With a prudent attitude, Bryan Ford \cite{2} proposed several less-understood issues which may emerge in the future. One is the availability issue. As the scale and complexity of clouds ever-growing, availability issue becomes more and more troublesome. Machida, etl \cite{5} studied the optimized software rejuvenation with live VM migration in clouds leveraging stochastic reward net. To explore the characteristics of system dependability in clouds especially the relationships between system failures and performance metrics, Guan set up a dependability analysis framework with a failures-metric DAGs (directed acyclic graph) in his paper \cite{6}. In industrial community, many cloud providers make a contract with service providers for the availability guarantee in SLA. Take Amazon EC2 for example, they provide 99.95\% availability in their SLA contract. But in our opinion, it's far from enough to only provide an availability number in SLA. Much more diverse availability items should be involved in SLA. Here we broaden the view of availability management in current cloud platform from three aspects: performance monitoring, availability analysis and availability recovery.

 Availability is traditionally defined as the `proportion of time a system is in a functioning condition'. However in a commercial software platform, the availability is not given when the response time exceeds the preset threshold. Moreover many studies \cite{6} have shown that system availability problems are relevant to performance metrics. To proactively troubleshoot and recover the availability problems, it's necessary to make some detailed performance metrics instead of conventional metrics monitored and reported if possible. The monitoring covers application-related metrics, process-related, OS-related, and even hardware-related metrics. Here we collect these metrics to assist the availability analysis rather than `sell performance' because the performance in different applications is too specific to be guaranteed \cite{8}. Another extension is to enrich the availability analysis such as finding the root causes of availability issues, calculating the MTTF (mean time to failure) and MTTR (mean time to recovery) or forecasting the failure time, etc. The last but not the least is to provide availability recovery mechanism such as migrating the VMs with a minimum cost, reconfiguring several parameters or restarting applications in an optimized cycle, etc. Recently many new cloud service modes come to the fore beside the traditional `IaaS', `PaaS' and `SaaS' such as `Resource as a service' \cite{8}, `Big data as a service'\cite{9}, etc. Following the hotly-used terminology `XaaS', we propose a new kind of service: availability as a service (abbreviated as AaaS). `AaaS' could be regarded as an auxiliary management mechanism in the clouds rather than a standalone service. It can be realized in IaaS, PaaS and SaaS although there may be some differences. In this paper we will stay focused on the implementation in IaaS. The availability information is delivered to cloud users in a service manner which is flexible and straightforward. `AaaS' enables the cloud users to get a deep profile of their applications and further repair the defects in the source code and reduce the physical resource utilization. Meanwhile, `AaaS' enables the cloud providers to attract more cloud users with a good reputation by reducing application failures. Thus `AaaS' is a win-win concept for cloud users and providers.

   According to the demonstration of `AaaS' above, we design and develop a prototype named `EagleEye'. `EagleEye' mainly contains four components including a management front-end, monitor agent, analysis engine and maintenance engine. The detailed description of these four components will be illustrated in Section 5. To make deep availability analysis, we integrated several data-driven and model-driven approaches which could be subscribed by cloud users. The development is still going and we will add more approaches in the future. 
\section{Motivations of AaaS} 

 As a common sense, management is a central characteristic of cloud computing. Much daunting work like setting up infrastructure, deploying applications and monitoring the runtime availability is largely simplified for cloud users. But the insufficiency of current management, the new requirements from customers and the new trend of IT technology determine new features should be involved in the clouds. We will demonstrate the detailed motivations in the following. \\
   $ \bullet$ \textbf{Insufficient responsibility assignment in SLA}. The commercial cloud providers are mainly responsible for the problems happened on their own cloud. However, who should take the responsibility for availability degradation or application failures? The answer is ambiguous in current responsibility assignment items. Under the attraction of cloud providers' ambitious advertisements, excessive trust from cloud users is put on the cloud computing. They may deploy their own applications without sufficient tests which buries a bomb for application failures. Therefore some failures are attributed to cloud users, not to the cloud operators. It's necessary to give an objective and acceptable criteria to distinguish the responsibility. This motivates us to make deep analysis of the root causes of availability issues. And assigning the responsibility according to the root causes.\\
$ \bullet$	 \textbf{Super privileges of the cloud providers}. The cloud providers have super privileges to access the whole infrastructure especial I/O activity information. Under the cooperation of users, they are even able to get the application-related metrics. Therefore cloud providers have great opportunity to find the root causes of availability issues. While due to the fact that cloud users have no visibility into the cloud and have to consult the cloud providers. As mentioned in paper \cite{4}, cloud providers should be involved in troubleshooting. In this case, why don’t we put the availability troubleshooting forwards to the cloud providers? \\
  $ \bullet$ \textbf{Defects of current availability management}. In current cloud platforms, they may provide several straightforward availability management mechanisms. Take Amazon EC2 for examples, they provide a real-time monitoring function named `CloudWatch', a elastic resource provisioning and releasing function named `AutoScaling' and a fault tolerance function named `Elastic LoadBalancing'. However they don't make deep analysis such as health measurement, failure prediction which is very important for auto scaling and load balancing. New availability approaches should be involved in the clouds.\\
  $ \bullet$ \textbf {The ever-growing complexity of application failures}. The cloud-based or cloud-centric applications become too sophisticated to understand the failure characteristics for cloud users. The transparent or non-transparent dependent relationships always hide the truth of application failures. Discovering the truth is not easier than finding a needle in hay. Hundreds of dimensions data will be considered. Real-time storing and processing these data are big challenges for cloud users. But it is relatively easy for cloud providers to fulfill these jobs by utilizing the large amount of idle physical resources. New IT technologies like Big Data \cite{9}, MapReduce\cite{11} can be implemented to accelerate the analysis. 


%

\section{Benefit from AaaS}
 'AaaS' is a win-win concept for cloud users and providers. Both of them can benefit and profit from this new service. In the following, we will demonstrate the benefit from the users' and providers' perspectives respectively. 

 \textbf{ From the cloud providers' perspectives, they can receive the following benefit:}

 a. Distinguish the responsibility for availability issues. Just as mentioned in the motivations, a relatively fair responsibility assignment principle could be given according to the root causes of the availability issues. Comparing to the situation that cloud providers take all the responsibility for availability issues, they can save some unnecessary cost.  

 b. Increase the utilization of shared resources. For high availability, the cloud providers usually leverage three strategies: redundancy, auto scaling and load balance. These strategies consume large amount of physical resources leading to low resource utilization and even worse they are ineffective. Based on the deep analysis in `AaaS', new strategies with low physical resource consumption can be raised such as reconfigure parameters or backup with a low redundancy.  

 c. Attract more cloud users. People are willing to spend money on high quality stuff. By supporting `AaaS', the cloud providers can attract more users to migrate their applications to the clouds. This makes the cloud providers more competitive in the cloud market. 

 d. Provide differentiated availability services. For mission-critical or safety-critical applications, they can support a high availability (e.g. 99.999\%) with a high price; but for those applications without so high requirements, they can support a relatively low availability (e.g.99.9\%) with a low price which brings a flexible choice for cloud users. 

\textbf{From the cloud users' perspectives, they can get the following benefit:}

  a. Enhance the experience of ultimate end users. By subscribing the services in `AaaS', the cloud users keep their applications at a high-level availability leading to better experience for end users. Driven by the good experience, the total revenues of cloud users will increase also.    
    
  b. Assist the understanding of applications. With the deep analysis in `AaaS', the cloud users can get some hidden defects in their applications' source code which may be hard to debug in the development. Through repairing the defects, it's possible to reduce the physical resource consumption or increase the performance of applications.
%

\section{Challenges and Opportunities of AaaS  }

Although `AaaS' brings some insightful ideas and beautiful envisions, it is not easy to be realized. In our opinion, the challenges can be attributed to three questions: what to monitor, how to price and what to analyze. 

   \textbf{Monitoring}. To get a deep availability analysis of the running applications, a detailed and holistic monitor framework is very necessary. The monitoring should cover the application-related, process-related, OS-related and even hardware-related metrics and include coarse-grained and fine-grained metrics. This requires more than hundreds of probes to be deployed in the infrastructure and maybe more than one monitor agents co-work on the same system. More difficult is that cloud providers should negotiate with cloud users to monitor the application-related metrics due to the privacy and security concerns. Luckily many excellent open source tools such as sysstat, collectd, perf can satisfy the primary requirements. However, other monitor tools used to collect the application-related metrics may be developed by cloud users or cloud providers.

 \textbf {Pricing}. A reasonable and acceptable pricing mechanism is essential for service selling. The price is influenced by many factors. According to our humble knowledge, we list several factors that should be concerned when pricing this service. First, the physical resources mainly CPU, memory and disk the service consumes. Second, the complexity of the analysis approaches. Third, the execution frequency. Fourth, the effectiveness of the availability recovery approaches. Currently, we can refer to the pricing approach of `CloudWatch' in Amazon EC2. For continually running services, a `pay-as-you-go' mode is given, for example set `4.5\$/month' with the frequency: 1/minute. For the other services, a `pay-per-use' mode may be appropriate. Apparently, new pricing approach will be proposed with the market growing. 

   \textbf {Analyzing}. To provide some data-driven and model-driven availability analysis approaches is the core of `AaaS'. However, what approaches should be integrated in `AaaS' and how to realize them? After reviewing a number of papers from 1990 to 2012 on availability in software systems, we propose several functions the `AaaS' may support: change or anomaly detection, correlation analysis, causality analysis, failure modeling, optimal decision analysis, etc. As to realization, we recommend R language to code these approaches due to the numerous libraries in the community. And then implement a high-level language such as Python or Java to call R methods.
\section{Design and Implementation of AaaS}
According to the demonstration of `AaaS' above, we design and develop a prototype named `EagleEye'. In this section, we will sketch the design, give an umbrella of the implementation and depict two analysis cases of `EagleEye'. 
\subsection{Design}
`EagleEye' mainly has four components including monitoring agent, management front end, analysis engine and maintenance engine (see Figure 1). Management front end is located in cloud user’s VM instance. It provides GUI and command line interfaces for users to interact with `EagleEye'. Through these interfaces, users can browser the availability information and performance information in real time; investigate the root causes of performance or availability anomaly or application failures; send control parameters such as the cycle of maintenance; subscribe services in `EagleEye'. Monitoring agent runs in Domain 0. It collects the runtime information from all the probes deployed in VM instances, hypervisor and hardware. The collected raw information can be stored locally or sent to the remote server according to the need. Analysis engine can be deployed in a fixed server or an idle server. The runtime information will be fed into this engine to be processed with aim-specific approaches such as failure modeling. A shining point in the engine design is we implement a `method bus' which is flexible for new analysis method to be merged in. The design methodology of maintenance engine is the same as analysis engine. This engine aims to determine an optimal availability recovery decision. We use XML-RPC protocol to transport control and availability information between management front end and the other two engines; use XML to pass message from analysis engine to maintenance engine; use text-file and socket format to share data between monitor agent and analysis engine. With this flexible design, new features can be easily integrated. \\
\begin {figure}
\psfig{figure=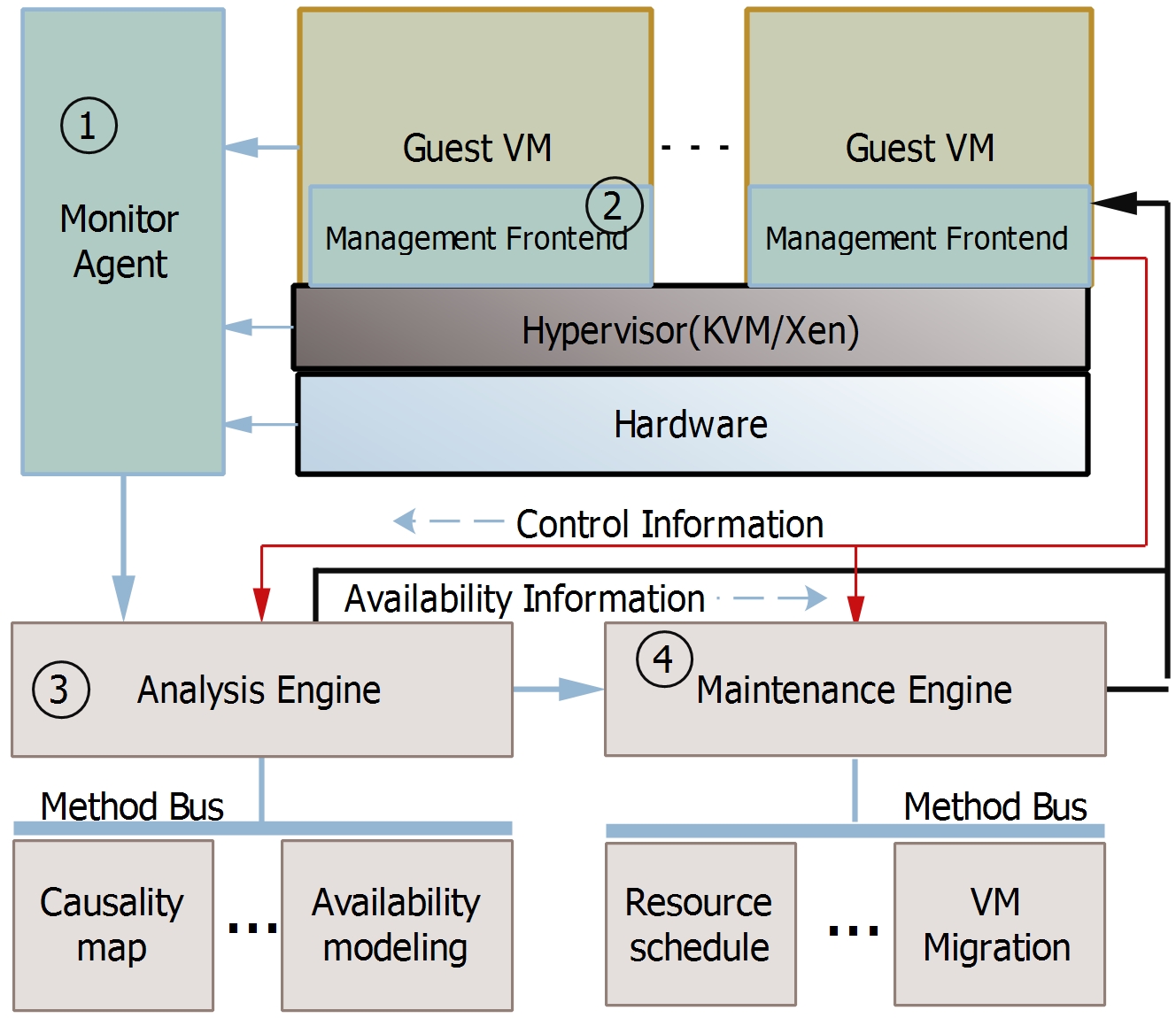,width=8cm}
\caption{The Framework of `EagleEye'}
\end{figure}
\subsection{Implementation}
The main body of `EagleEye' is coded by python language which is an almighty glue language to unify other logical components developed with other languages. To get a comprehensive profile of the running applications, we integrated multiple off-the-shelf tools and self-developed tools. To get the fine-grained information like system call count or latency, we write several scripts leveraging SystemTap \cite{12} developed by IBM based on kernel instrumentation. All the analysis and decision approaches are developed with R language. Currently, this prototype is deployed on a controlled environment which contains 8 VM instances. And RuBiS, a multi-tiered application benchmark, is used to test the effectiveness. Here we only give two analysis cases in `EagleEye' due to the limited space.

    As stated in \cite{2}, deep resource dependency analysis is an effective method to resolve the hidden failure correlations. We build a two-level dependency graph to infer the root causes of availability issues. The high level dependency is used to locate the source of problem at service level (see figure 2). The low level is a performance metric dependency graph which is constructed by PC algorithm \cite{13} (see figure 3). Through traversing these two graphs, root causes can be located at fine granularity. According to the root causes, cloud providers can recover application availability with low cost and cloud users can check and repair the defects in their application source code.   

    Another case is to implement multi-scale entropy \cite{14} to measure the health state of running applications. According to the conclusion of paper \cite{14},  a higher entropy value implies a higher failure risk. Therefore it's easy to query the health state of the applications in real time. If the entropy exceeds the preset threshold, maintenance action will be triggered in order to recovery the availability. Figure 4 depicts the multi-scale entropy variations of RuBiS application with time increasing. From figure 4, we can see the application steps slowly from health state to faliure state with entropy increasing from low to high.       
\begin {figure}
\psfig{figure=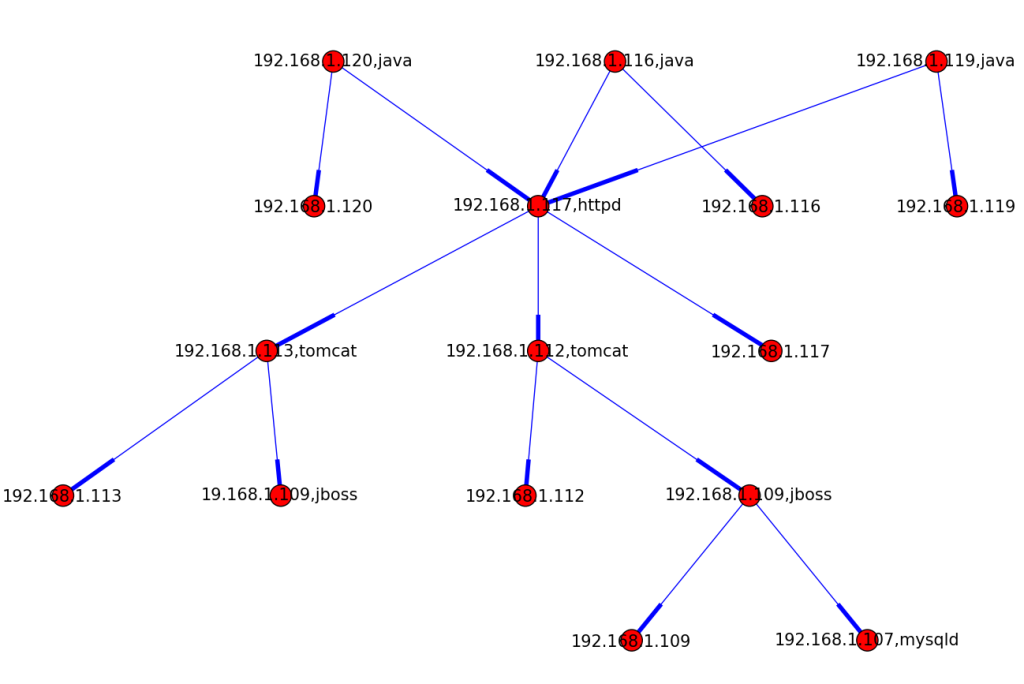,width=8cm}
\caption{The high-level service dependency.Every node is a tuple with two elments: IP address and service name.}
\end{figure}

\begin {figure}
\psfig{figure=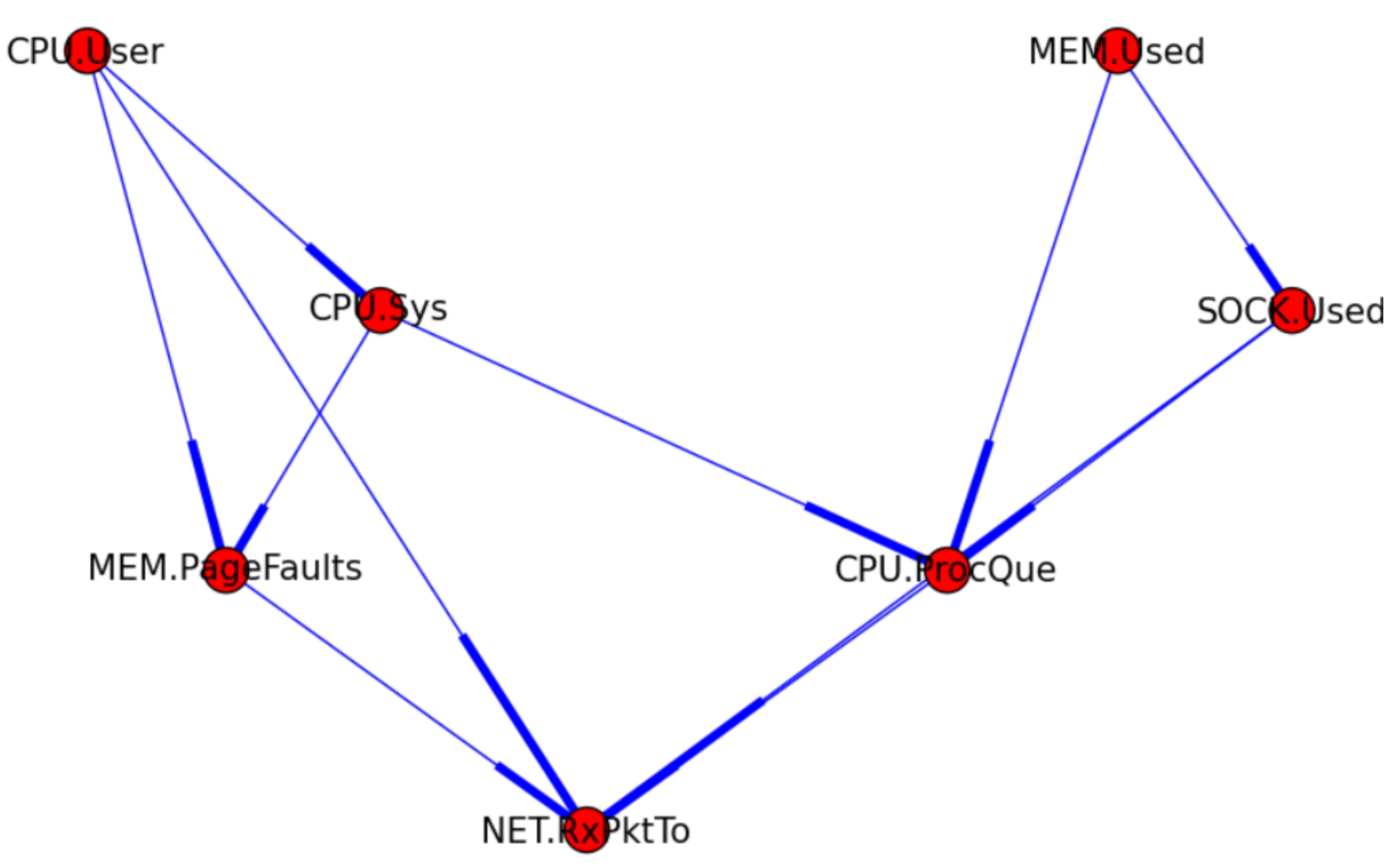,width=8cm}
\caption{The low-level metric dependency. Every node is a performance metric.}
\end{figure}

\begin {figure}
\psfig{figure=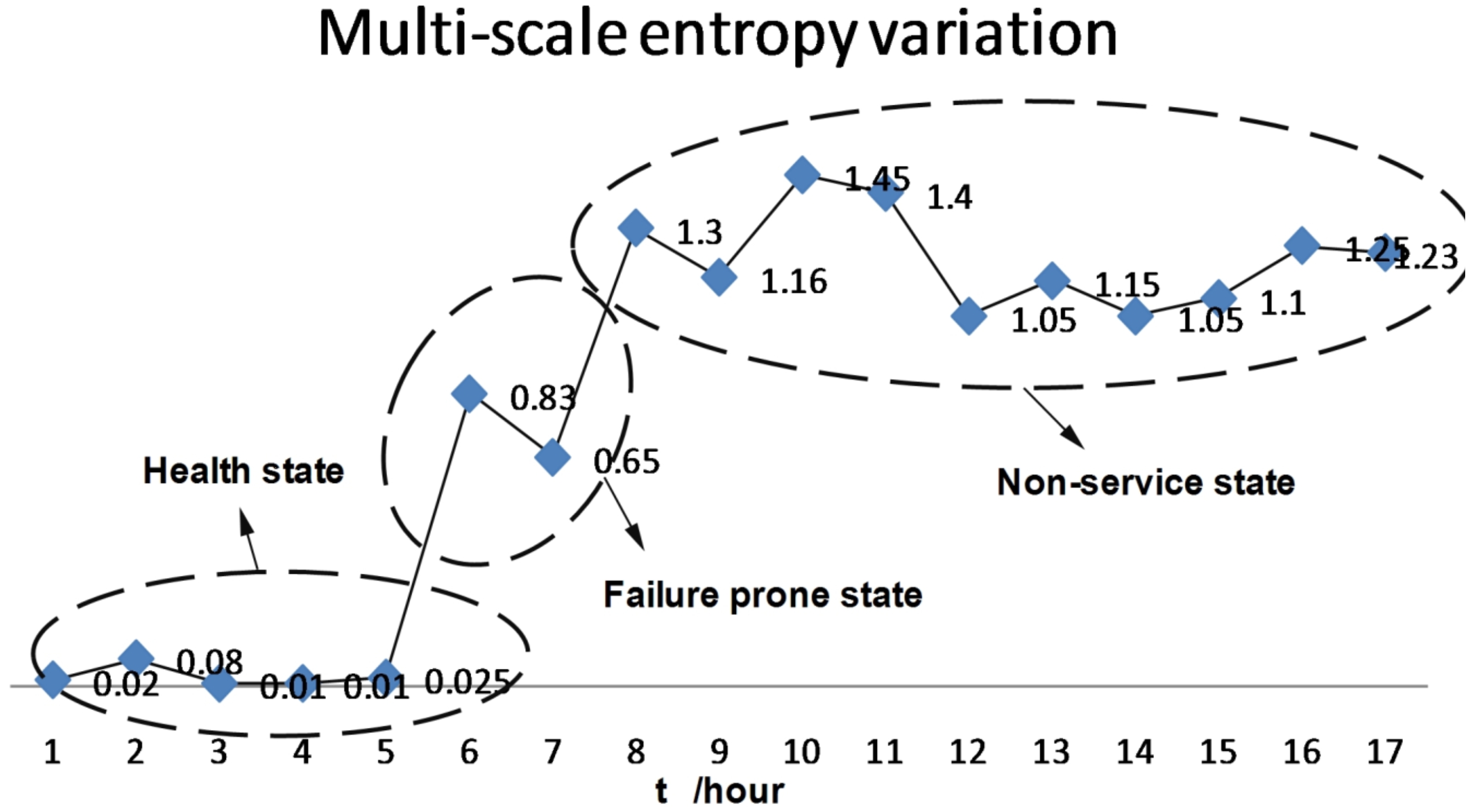,width=8cm}
\caption{The multi-scale entropy variation}
\end{figure}

\section {Conclusion and Future Work}
 Considering the importance of availability in cloud computing, we propose a new kind of service named `AaaS'. Cloud users and cloud providers can both benefit and profit from `AaaS'.  Through designing and developping a prototype `EagleEye', we make the `AaaS' a reality. And it is hopeful to merge `AaaS' into the current availability managment mechanism of cloud computing. We will continue to add more analysis approaches such as condition-based maintenance based on PHM (i.e. proportional hazard model) to the prototype. 
\section{Acknowledgments}
Thanks to the all the members in our research regroup. This research is sponsored by NSFC with Grant No. 60933003
%
%
%
%
{\footnotesize \bibliographystyle{acm}
\bibliography{template}}
\end{document}